\documentclass[preprint2]{aastex62}
\usepackage[utf8]{inputenc}
\usepackage{graphicx}
\usepackage{wrapfig}
\usepackage{array}
\usepackage{tabularx}
\usepackage{lineno}
\usepackage{amsmath}
\usepackage{xcolor}
\begin{document}

\title{Uranus' hidden narrow rings}
\author{M.M. Hedman}
\affil{Department of Physics, University of Idaho, Moscow Idaho 83843}
\author{R.O. Chancia$^{1,}$}
\affil{Chester F. Carlson Center for Imaging Science, Rochester Institute of Technology, Rochester NY 14623 \\ }
%\date{\today}

\begin{abstract}
In addition to its suite of narrow dense rings, Uranus is surrounded by an extremely complex system of dusty rings that were most clearly seen by the Voyager spacecraft after it flew past the planet. A new analysis of the highest resolution images of these dusty rings reveals that a number of them are less than 20 km wide. The extreme narrowness of these rings, along with the fact that most of them do not appear to fall close to known satellite resonances, should provide new insights into  the forces responsible for sculpting the Uranian ring system.

\vspace{0.4in}

\end{abstract}

\section{Introduction}

Uranus' ring system is dominated by a series of nine dense, narrow rings discovered in 1977 and now designated 6, 5, 4 $\alpha$, $\beta$, $\eta$, $\gamma$, $\delta$ and $\epsilon$ \citep{Elliot77, Millis77, BK77}. While the $\epsilon$ ring ranges between 20 and 100 km in width, the other rings are all 2-10 km wide \citep{French86, French91, Nicholson18}. This collection of dense, narrow rings is very different from Saturn's dense broad rings and the dusty rings surrounding Jupiter and Neptune. Indeed, the closest analogs to the Uranian rings are the dense, narrow rings around small bodies like Chariklo and Haumea \citep{BragaRibas14, Ortiz17}, and it is still not clear why Uranus' rings are so narrow, nor why these narrow rings are located where they are. 

\begin{figure}
\resizebox{3.in}{!}{\includegraphics{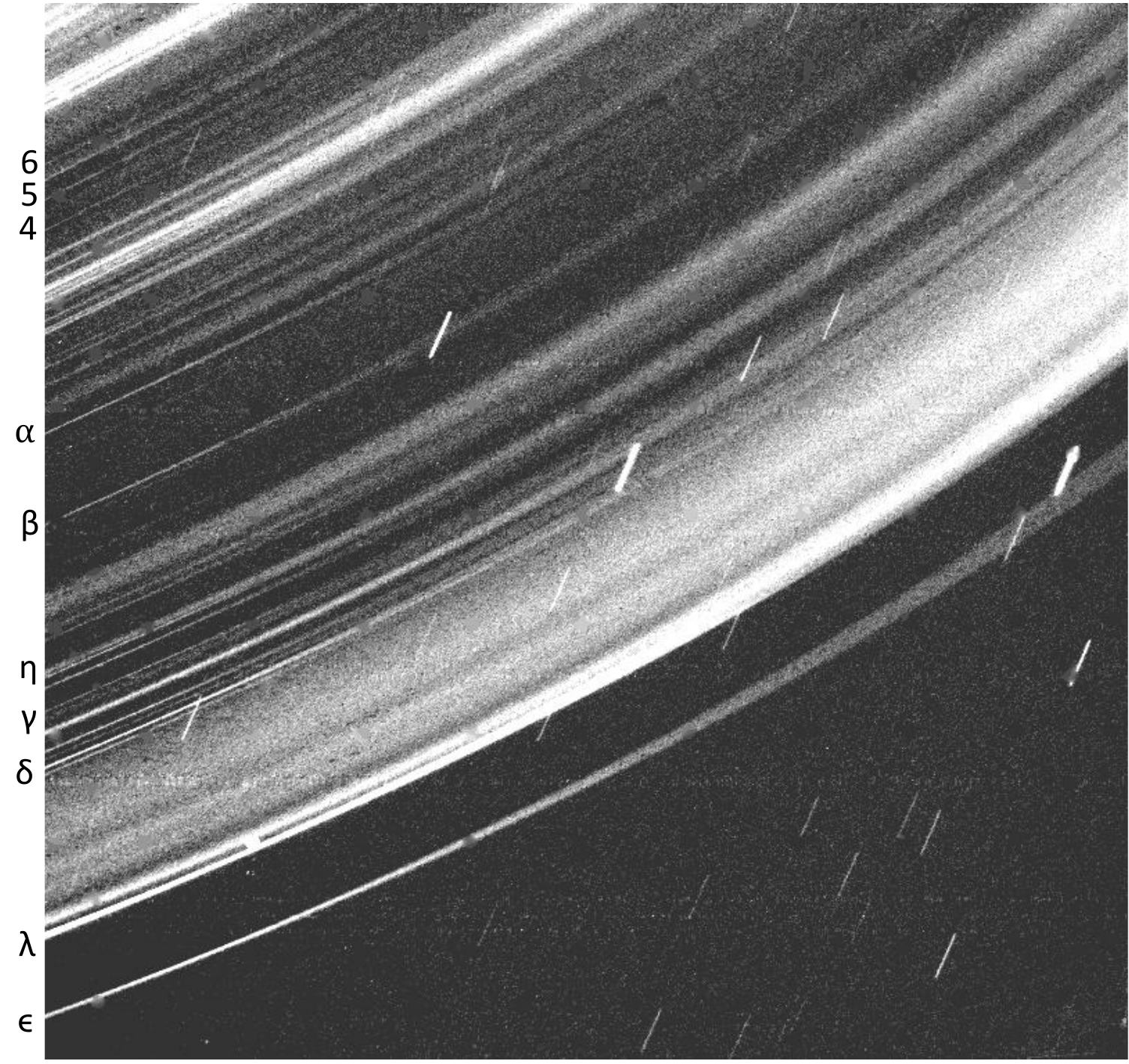}}
\caption{Voyager 2 Wide Angle Camera image C2685219  of the Uranian ring system obtained while the spacecraft flew through Uranus' shadow (Planetary Photojournal image PIA00142, with annotations showing locations of named rings). Note that the image becomes increasingly smeared from left to right, and ring radius increases from top to bottom. Also note that the bright feature just interior to the $\lambda$ ring is an image artifact caused by automatic algorithms filling a reseau mark.}
\label{wac}
\end{figure}

 \begin{figure*}
\resizebox{6.5in}{!}{\includegraphics{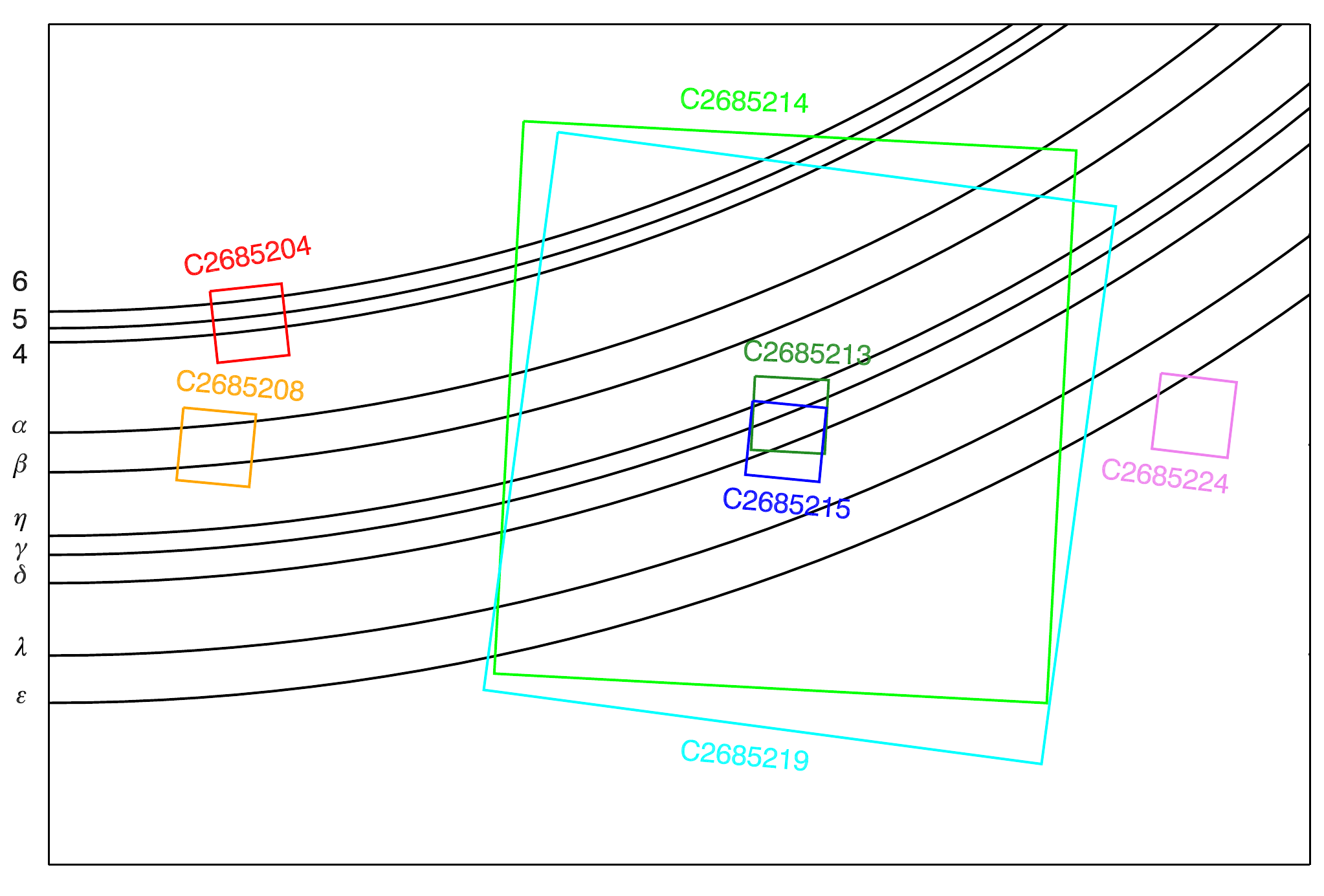}}
\caption{Map showing the locations observed by the seven images obtained during the Voyager 2 high-phase observing sequence.}
\label{map}
\end{figure*}

\begin{table*}
\caption{Geometrical Parameters for the Voyager Images}
\label{ims}
\hspace{-.7in}
\resizebox{7.5in}{!}{\begin{tabular}{cccccccc}\hline
Image Name & Camera &  UTC & Exposure & Phase & Emission & Longitude & Image Resolution  \\ 
&  & &Time (s) & Angle (deg) & Angle (deg) & (deg) & (km)$^a$  \\ \hline
C2685204 & NAC & 1986-024T21:21:33 & 5.76 & 172.7 & 177.9 & 263.5 & 3.95 \\ 
C2685208 & NAC & 1986-024T21:24:45 & 5.76 & 172.0 & 177.4 & 264.9 & 3.99\\ 
C2685213 & NAC & 1986-024T21:28:45 & 5.76 & 172.4 & 179.5 & 248.3 & 4.05\\
C2685214 & WAC & 1986-024T21:28:45 & 1.92 & 172.5 & 179.5 & 248.4 & 31.5\\
C2685215 & NAC & 1986-024T21:30:21 & 5.76 & 172.4 & 179.7 & 248.7 & 4.07 \\
C2685219 & WAC & 1986-024T21:33:33 & 96.0 & 172.4 & 179.6 & 248.4 &  32.0\\ 
C2685224 & NAC & 1986-024T21:37:33 & 5.76 & 172.4 & 178.4 & 238.4 & 4.18 \\\hline
\end{tabular}}

$^a$ Nominal image resolution assuming resolutions of 18 $\mu$rad for the NAC and 140 $\mu$rad for the WAC (numbers derived from PDS instrument description files inst\_cat\_na1.txt and inst\_cat\_wa1.txt).

\vspace{.5in}
\end{table*}

The Voyager-2 spacecraft's encounter with Uranus revealed that in addition to the 9 dense rings, Uranus also had a number of dusty rings.  The most dramatic picture of these rings was Wide-Angle Camera (WAC) image C2685219\footnote{Note the image designations used here correspond to the filenames provided for the PDS archive. In earlier works  \citep[e.g.][]{Esposito91} this image was designated FDS 26852.19.}, obtained at a phase angle of around 172.4$^\circ$ after the spacecraft flew past the planet \citep{Smith86, Ockert87, MT90, Esposito91}. This image showed a complex array of ring structures on a wide range of scales (see Figure~\ref{wac}). The brightest feature in this image was the $\lambda$ ring between the $\delta$ and $\epsilon$ rings. The $\lambda$-ring's brightness at high phase angles indicates that it is composed of dust-sized particles, unlike the other named rings. However,  Voyager occultation measurements also revealed that this ring was also only a couple of kilometers wide \citep{Holberg87, Colwell90, Showalter95}, indicating that both dense and dusty Uranian rings could be very narrow.

C2685219 was not the only image that was obtained at these very high phase angles. There was a second Wide-Angle Camera image (C2685214) that had a shorter exposure and so only clearly showed the $\lambda$ ring  \citep{Showalter95}. Furthermore, there was a series of five Narrow-Angle Camera (NAC) images obtained during this time, each one targeted at the locations of one or more of the dense rings (see Figure~\ref{map} and Table~\ref{ims} for summaries of the image geometries). These images have never been properly analyzed \citep{Ockert87}, most likely because upon initial inspection they look largely featureless. However, closer inspection reveals that these five images do contain signals from the rings. In fact, these five narrow-angle camera images are the highest-resolution images of the Uranian rings, and they reveal that the $\lambda$ ring is not unique in being an extremely narrow dusty ring.  

This paper analyzes these Narrow-Angle Camera images to obtain information about the properties of all the narrow ring features they contain. Section~\ref{methods} describes how the images were processed to obtain high-pass filtered radial brightness profiles. Section~\ref{results} discusses the narrow features identified in these profiles, which includes several dusty ringlets that are less than 10 km in width. Finally, Section~\ref{discussion} discusses the potential implications of these narrow dusty features for the Uranian ring system.

\begin{figure}
\resizebox{3.2in}{!}{\includegraphics{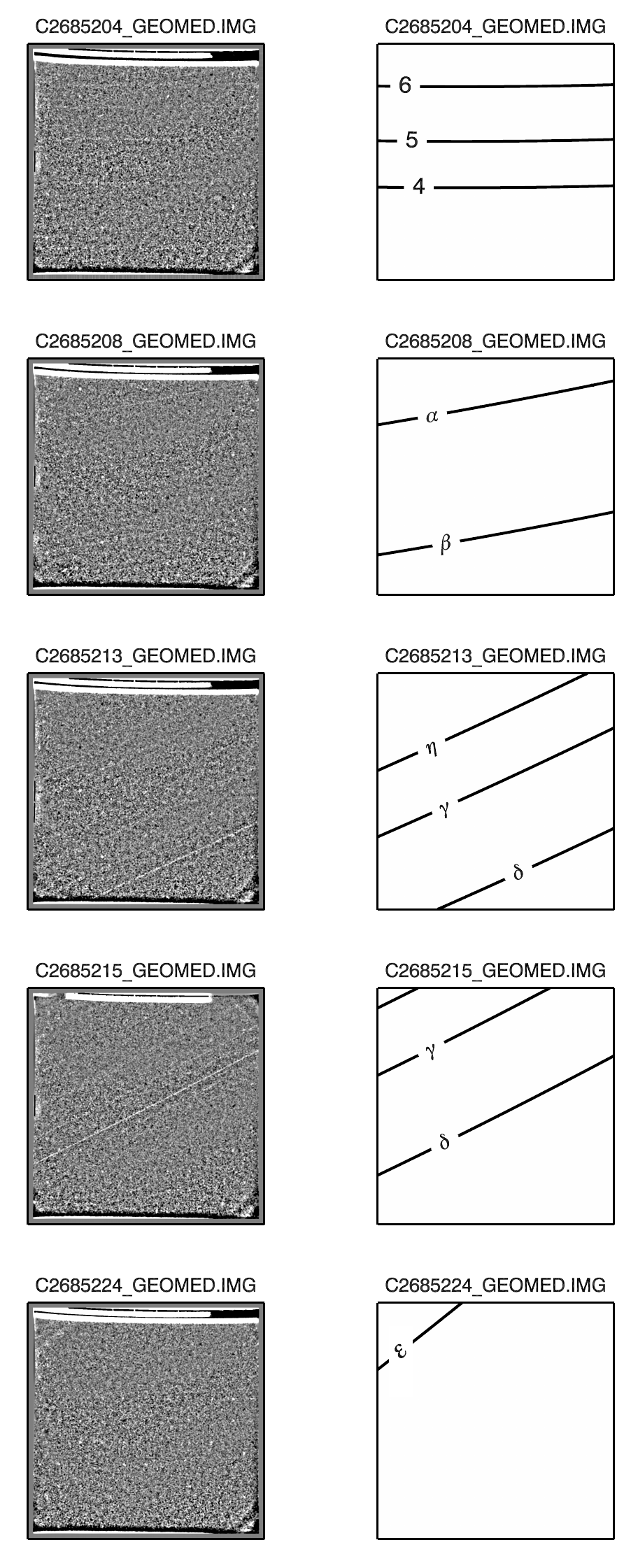}}
\caption{Narrow-Angle Camera images of the Uranian ring system. On the left are geometrically corrected, spatially filtered versions of the 5 images, while on the right are the predicted locations of the named rings in each frame. Hints of both the named rings and new narrow ring structures can be seen in the images.}
\label{nac}
\end{figure}

\begin{table}
\caption{SPICE kernels used for this analysis.}
\label{kernels}
{\begin{tabular}{l}
vg2\_v02.tf \\
vg2\_issna\_v02.ti \\
vg2\_isswa\_v01.ti \\
naif0010.tls \\
vg200022.tsc \\
pck00010.tpc \\
ura091-rocks-merge.bsp \\
vgr2\_ura083.bsp \\
ura111.bsp \\
vgr2\_super.bc \\
vg2\_ura\_version1\_type1\_iss\_sedr.bc \\
\end{tabular}}
\end{table}

\section{Methods}
\label{methods}

This analysis uses calibrated and geometrically-corrected versions of the images listed in Table~\ref{ims} archived on the Ring-Moon Systems node of the Planetary Data System. Each of these processed images consists of a square array of 1000$\times$1000 pixels, which are upsampled from the original 800$\times$800 pixel images \citep{Smith77} and include corrections for geometric distortions across the field of view. These images were geometrically navigated using the SPICE kernels \citep{Acton96} listed in Table~\ref{kernels}.  Note that the available kernels specifying the camera pointing (vg2\_ura\_version1\_type1\_iss\_sedr.bc) contain pointing errors amounting to a significant fraction of a Narrow-Angle Camera frame. The image geometry therefore needed to be corrected based on the observed positions of stars and known ring features. 

In practice, known ring features can only be seen in the Narrow-Angle Camera images  after subtracting a copy of the image that was boxcar-average smoothed with a smoothing length of 50 pixels. Figure~\ref{nac} shows the resulting high-pass filtered images (smoothed over $5\times5$ pixels for display purposes). These images  all contain linear features, some of which correspond to the named dense rings, while others are novel. Note that unlike the WAC image shown in Figure~\ref{wac}, there are no obvious changes in the sharpness of the narrow features across these NAC images, most likely because the much lower exposure times used for these images (see Table~\ref{ims}) greatly reduced image smear.

Image C2685204 was targeted at the 4, 5 and 6 rings. Faint horizontal lines can be seen around the expected locations of all three rings, and there may be another line roughly midway between the 5 and 6 rings.  Image C2685208 covered the region around the $\alpha$ and $\beta$ rings, and faint lines can be seen near the expected positions of both rings.  Images C2685213 and C2685215 cover overlapping parts of the region containing the $\eta$, $\gamma$ and $\delta$ rings. Both these images show multiple lines parallel to the expected trends for these rings, and two of these fall close to the expected location of the $\gamma$ and $\delta$ rings themselves. However, it is important to note that the brightest band in these images does {\bf not} fall at the expected location of these rings. It is instead located about 80 km interior to the $\delta$ ring. These two images therefore provide the most intriguing evidence that these Narrow-Angle Camera images contain information about previously unknown narrow rings. Finally, image C2685224 was targeted at the $\epsilon$ ring, and there is a band near the upper left corner of the image that is likely due to this ring.

The ring signals can be better quantified if we transform the raw imaging data into profiles showing the average brightness as a function of ring-plane radius. However, in order to do this averaging, the images need to be geometrically navigated. As mentioned above, the camera orientations encoded in the available kernels have significant errors and so each image's geometry needs to be corrected based on the observed locations of rings and stars. Only two images (C2685208 and C2685215) contained a star that could be used for this navigation (TYC 1296-591-1 and TYC 1282-163-1, respectively). Fortunately, this was sufficient to confirm the identities of the  $\alpha$ and $\beta$ rings in C2685208, and to demonstrate that two of the ringlets in  C2685215 were at positions consistent with the expected locations of the $\gamma$ and $\delta$ rings (as well as that the brightest feature in this image fell slightly interior to the $\delta$ ring). We could  therefore navigate image C2685213 based on the locations of those ring features. For image C2685204  the most prominent features had spacings consistent with those expected for the 4/5/6 rings, so those features were used for navigation. Finally, C2685224 was navigated so that the bright band  that was most likely the $\epsilon$ ring fell at the appropriate location. Note that all these adjustments required similar offsets relative to the nominal pointing kernel, which gives us further confidence that our navigation of these images is correct. 

After navigating the images, we can compute radial brightness profiles. Since we are interested in the narrow ring features here, we again apply a high-pass filter to the images by subtracting a version of the image that has been boxcar-average smoothed over 50 pixels. We verified that the shapes of the resultant peaks associated with the narrow ringlets are not particularly sensitive to the smoothing scale. We then exclude all pixels within 100 pixels of each edge of the frame to  remove areas affected both by the filter and the original frame boundaries within the geometrically-corrected images. We then use the appropriate SPICE kernels to compute the radius and longitude for each pixel and average over longitude to produce a profile of brightness versus radius, and we multiply these brightness values by the cosine of the emission angle $\mu$ to obtain values of ``Normal $I/F'$' or $\mu I/F$ for each radius. We also compute the statistical error on these brightness values based on the variance of the data within each radius bin. However, it is important to note that these uncertainties do not account for instrumental artifacts in the images, such as masked  reseau marks, stray light and signal variations between scanlines. The systematic errors in the brightness profiles generated by these phenomena are difficult to quantify {\it a priori} and can be particularly important near the edges of the images.

The positions of the peaks associated with the named rings in these profiles were generally within 10 km of their expected positions. We removed these residual offsets by simply shifting the radius scale of each profile so that the peak occurred at the location one would expect for a subset of the named ringlets based on standard models, including their eccentricities \citep{French91, French20}. For C2685204 we used the positions of the 5 and 6 rings, since the profile for the 4 ring was noticeably wider and asymmetric. For C2685208 we used the positions of the $\alpha$ and $\beta$ rings. For C2685213 and C2685215 we used the position of the $\gamma$ ring since the peak corresponding to the $\delta$ ring was also asymmetric.  Note that these corrections assume that the material visible here is the same as that observed in the occultations used to model the rings' positions. In principle, there could be offsets between the location of the dense rings observed in occultations and the brightest regions seen at high phase angles.  Such offsets could cause slight biases in the estimated feature locations, but these are unlikely to be more than a couple of kilometers.

To put these profiles in context, we also computed a complete profile of the ring system from the Wide-Angle Camera image C2685219, similar those found in prior work \citep{MT90, Esposito91}. We navigated the image based on the positions of known ring features, and only considered data in columns 30-70 of the  calibrated, geometrically-corrected image, where smearing was minimal. Again, these data were averaged over longitude and multiplied by the cosine of the emission angle to produce a radial brightness profile that has lower resolution but better signal-to-noise than the profiles derived from the NAC images. Finally, we remove a linear background trend from the WAC profile to make the small-scale brightness variations easier to see.

\begin{figure*}
\resizebox{6.5in}{!}{\includegraphics{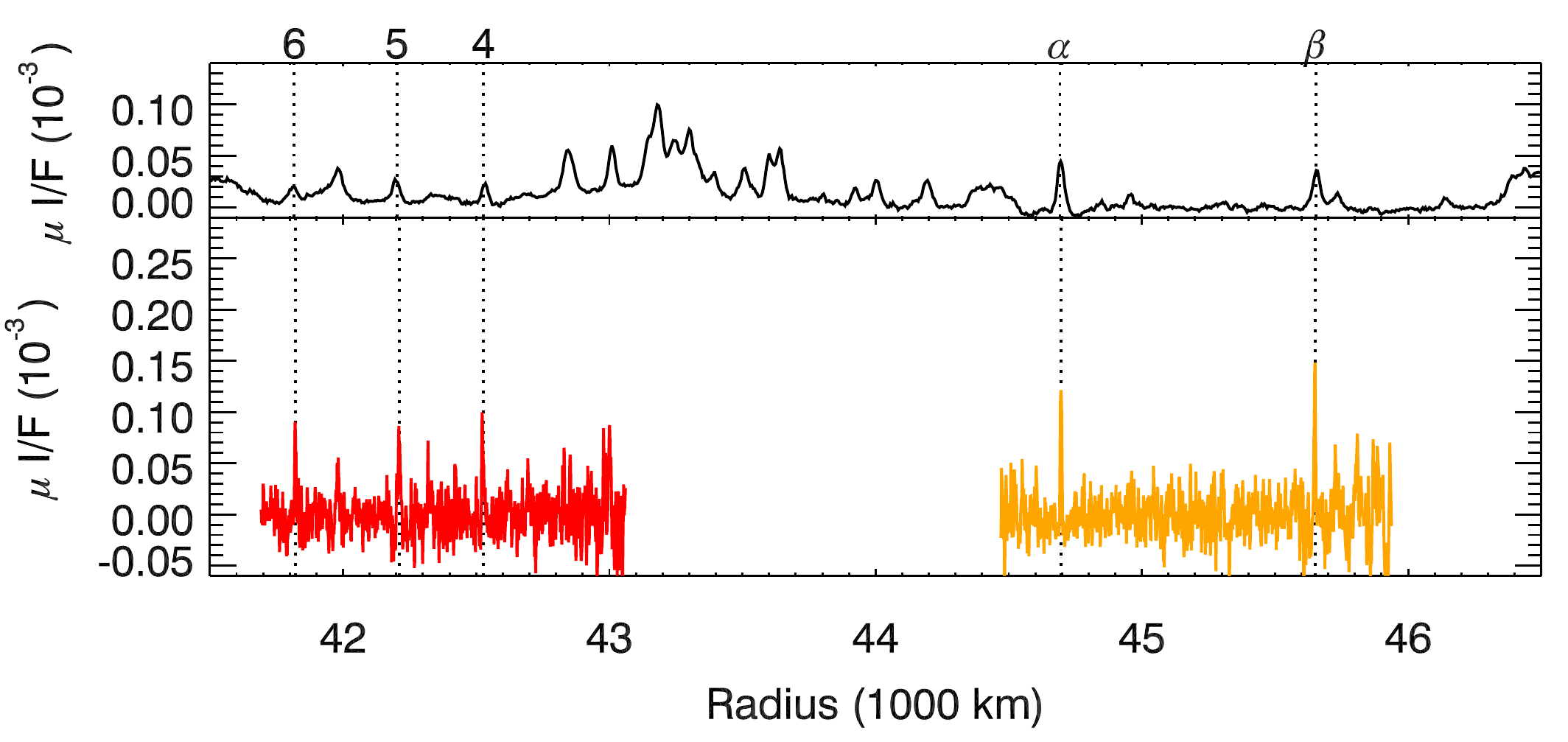}}
\resizebox{6.5in}{!}{\includegraphics{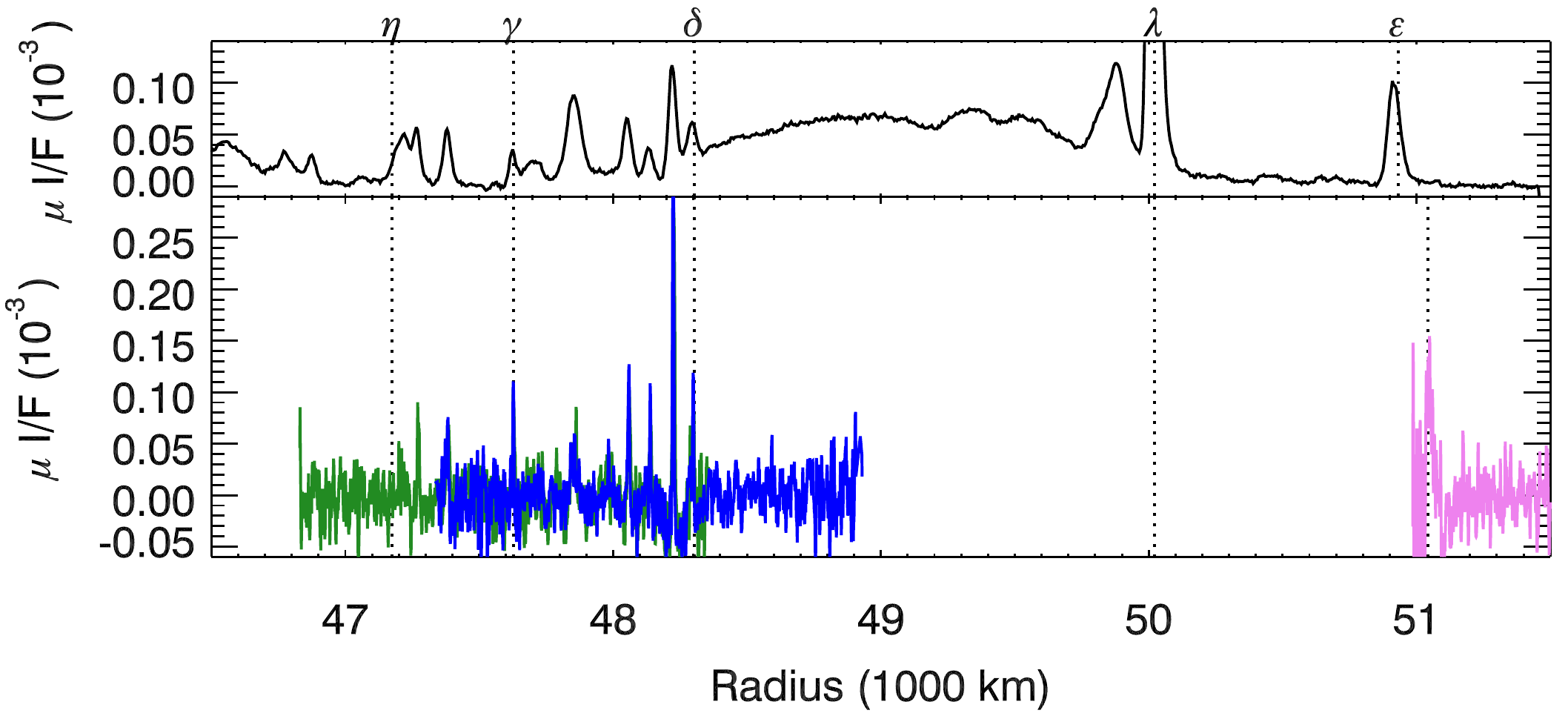}}
\caption{Overview of the radial brightness profiles derived from the NAC images (red=C2685204, orange=C2685208, green=C2685213, blue=C2685215, violet=C2685224), compared with a profile derived from the WAC image C2685219 in black. Vertical dashed lines mark the expected named ringlet positions based on current models}
\label{overview}
\end{figure*}

\section{Results}
\label{results}

\begin{figure*}
\resizebox{6.5in}{!}{\includegraphics{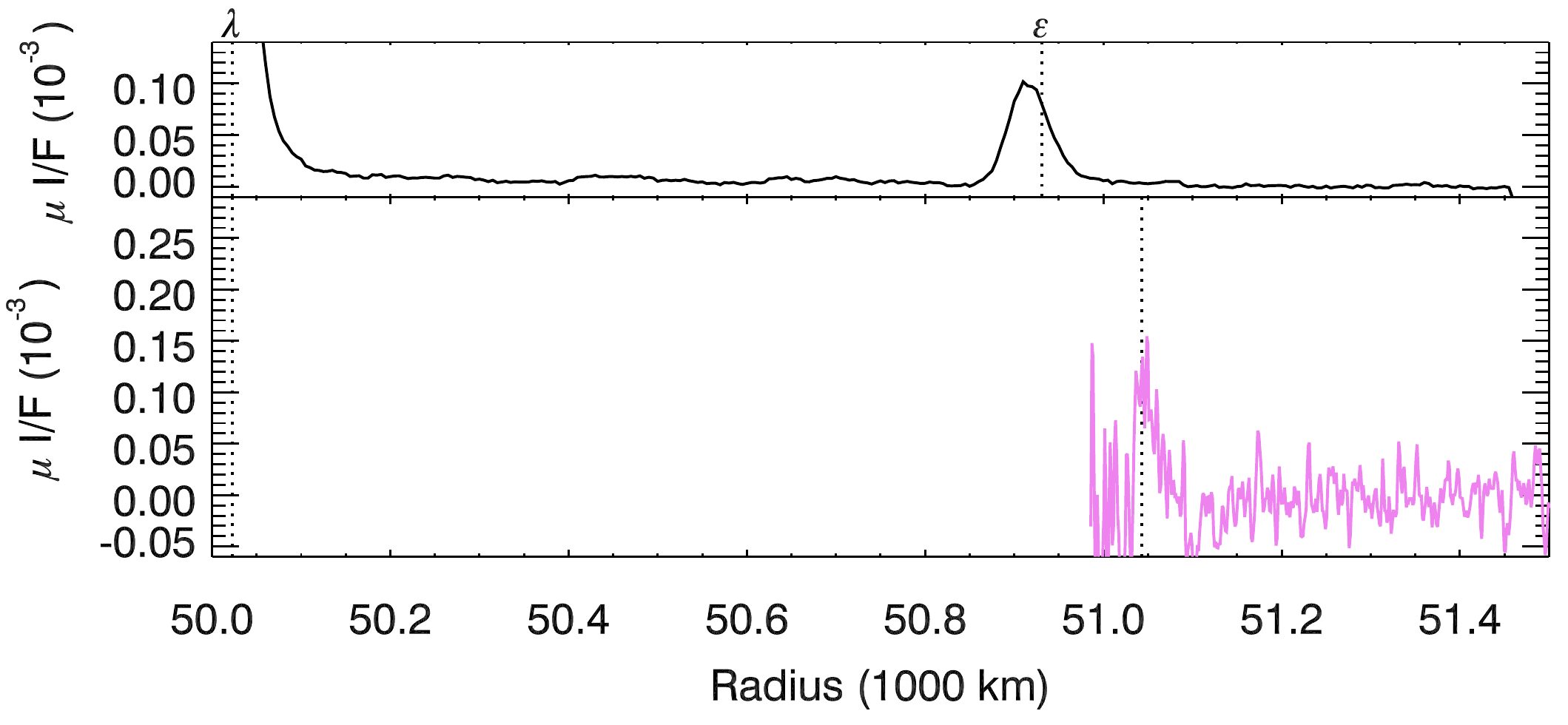}}
\caption{The radial brightness profile derived from the NAC image C2685224, compared with a profile derived from the WAC image C2685219. Vertical dashed lines mark the expected named ringlet positions based on current models. Note the predicted position of the $\epsilon$ ring changes significantly between the two images due to that ringlet's large eccentricity.  Both profiles show a single, relatively broad peak that can be attributed to the $\epsilon$ ring. }
\label{e}
\end{figure*}

\begin{figure*}
\resizebox{6.5in}{!}{\includegraphics{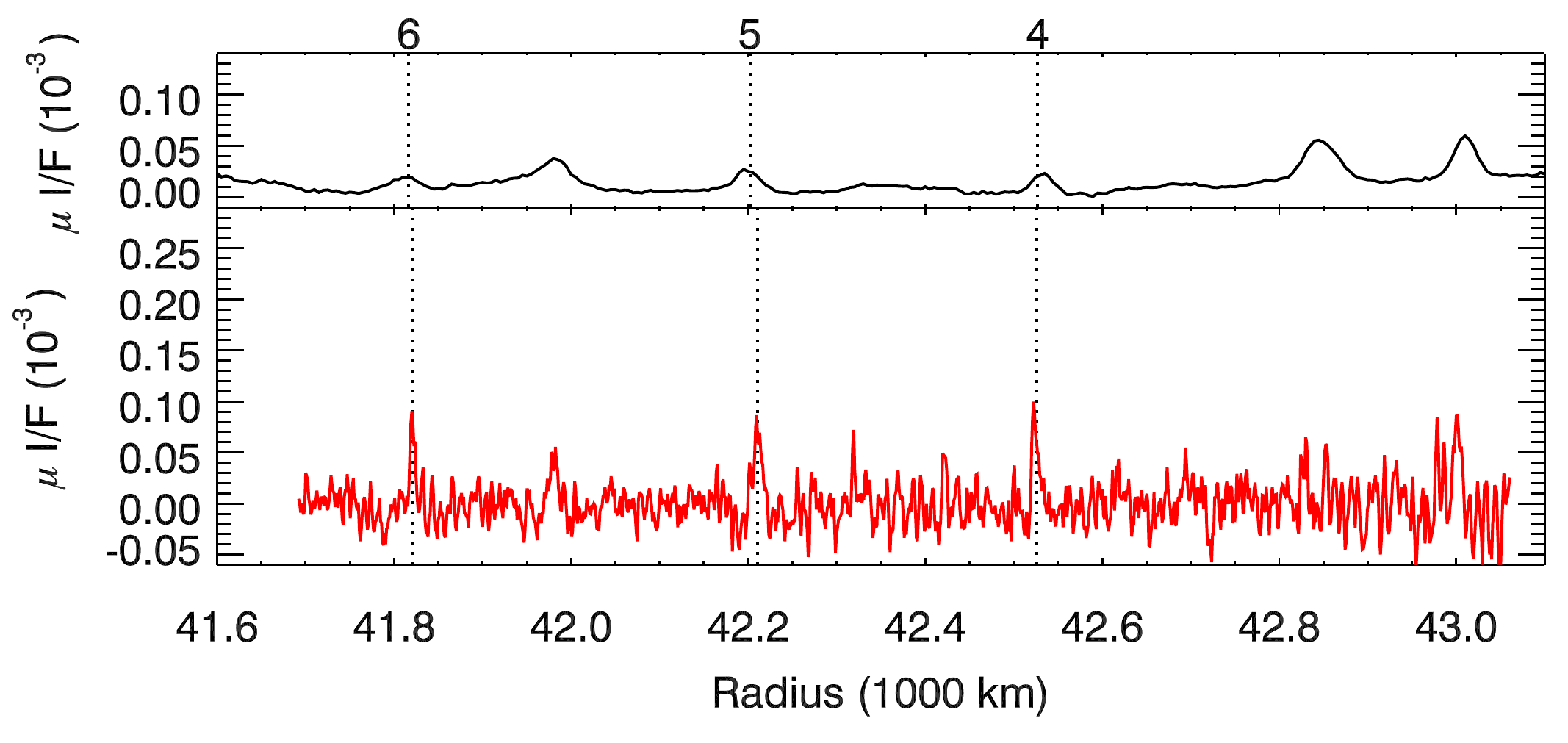}}
\caption{The radial brightness profile derived from the NAC image C2685204, compared with a profile derived from the WAC image C2685219. Vertical dashed lines mark the expected named ringlet positions based on current models. Note there are three peaks near the expected locations of the 4,5 and 6 rings, and there is another narrow ringlet between the 6 and 5 rings at 41,980 km. There may also be peaks around 42,830 km and 43,000 km corresponding to the peaks in the wide-angle camera profile, but the signal-to-noise is lower here. }
\label{456}
\end{figure*}

\begin{figure*}
\resizebox{6.5in}{!}{\includegraphics{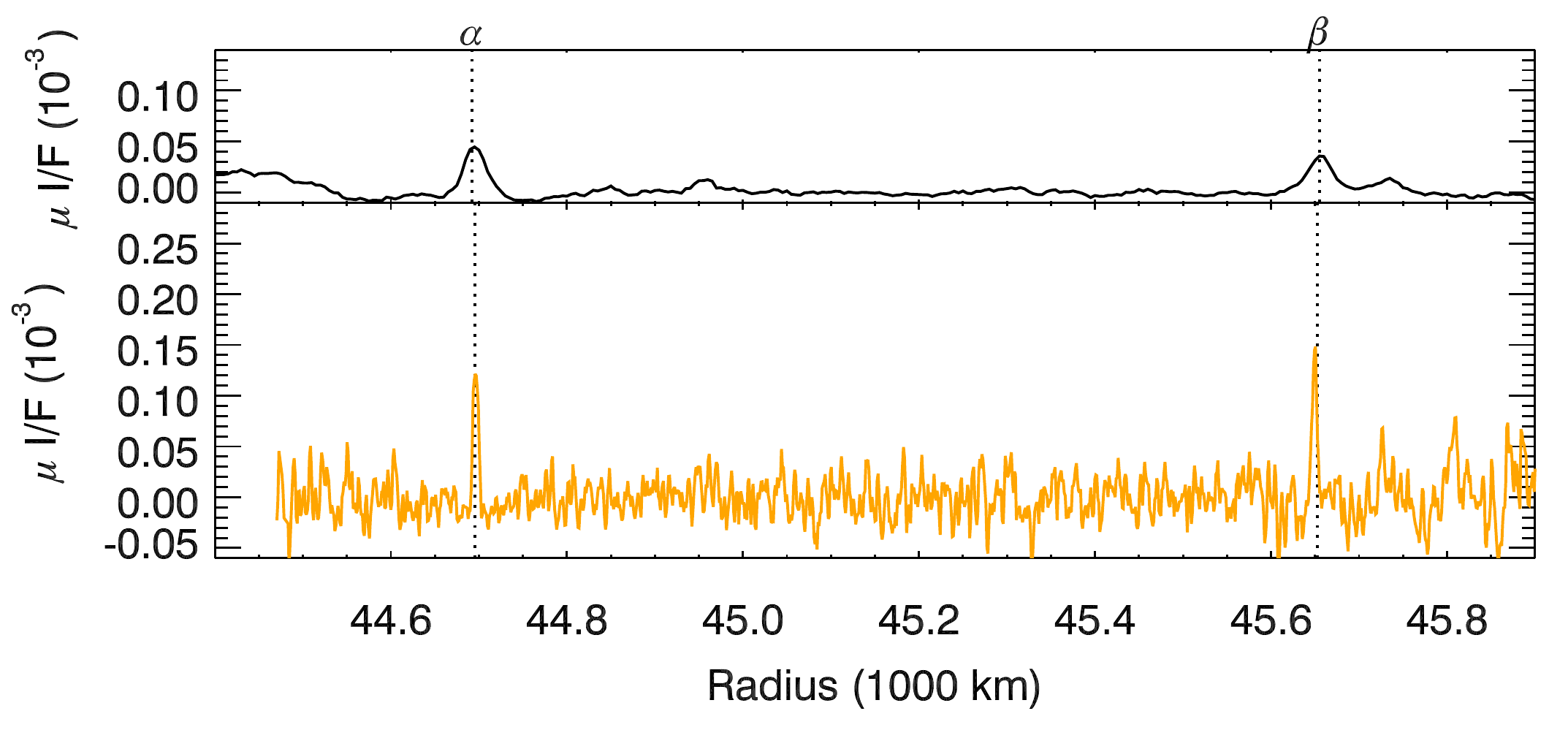}}
\caption{The radial brightness profile derived from the NAC image C2685208, compared with a profile derived from the WAC image C2685219. Vertical dashed lines mark the expected named ringlet positions based on current models. Note there are two peaks near the expected locations of the $\alpha$ and $\beta$ rings. There may also be a feature corresponding to a peak in the WAC profile just outside the $\beta$ ring, but there is a semi-periodic signal variation here that complicates detection of that signal. }
\label{ab}
\end{figure*}

\begin{figure*}
\resizebox{6.5in}{!}{\includegraphics{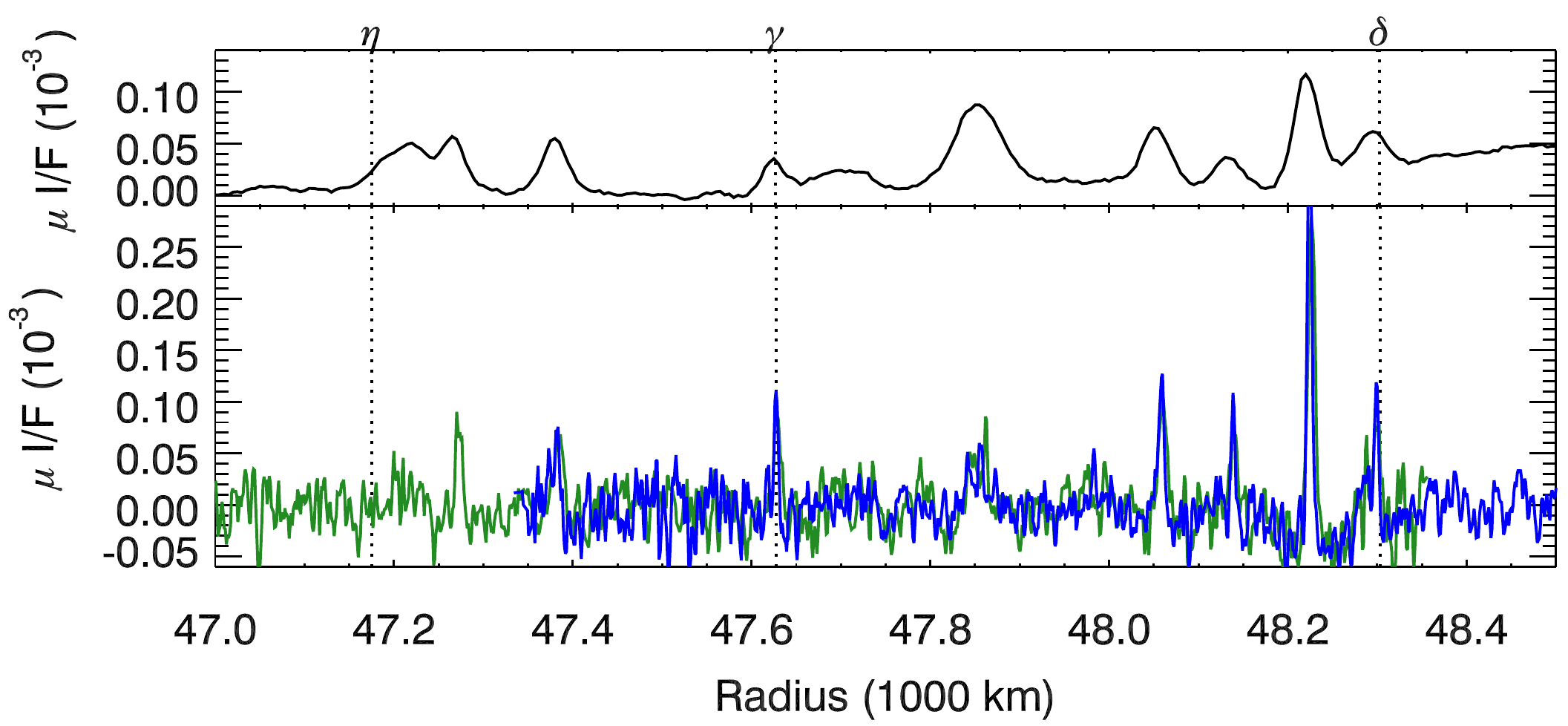}}
\caption{The radial brightness profiles derived from the NAC  images C2685213 (green) and C2685215 (blue), compared with a profile derived from the WAC image C2685219. There is no clear peak associated with the $\eta$ ring, but there are clear peaks associated with both the $\gamma$ and $\delta$ rings. In addition, there are narrow peaks at 47,270 km, 48,060 km, 48,140 km and 48,230 km, the last of which is brighter than any other feature captured in the NAC profiles. There are also somewhat wider peaks at 47,380 km and 47,860 km.}
\label{gd}
\end{figure*}

Figure~\ref{overview} provides an overview of the profiles derived from the high-phase NAC images, while Figures~\ref{e}-\ref{gd} provide close-ups of interesting regions. Recall that the NAC profiles are high-pass filtered and so are insensitive to broad ring features like the sheet of dust seen in the WAC profile between the $\delta$ and $\lambda$ rings. Nevertheless, there are peaks in the NAC profiles that are correlated with the peaks in the WAC profile, and are generally much narrower, thus demonstrating that many of the features in the WAC profile are unresolved. 

Meanwhile, Table~\ref{params} provides summary statistics for the identifiable ring signals in these profiles. These parameters are computed by finding the radius of the peak brightness and then fitting the region within 50 km of this peak location to a model of a Gaussian peak plus constant background. We subtract the constant background level from this fit to determine the peak normal $I/F$ and we estimate the Gaussian FWHM of the peak from the fit parameters. Finally, we integrate the brightness above background over a region $\pm50$ km around the peak to compute the normal equivalent width, or $NEW$ of the peak. Note that while it is difficult to assign errors {\sl a priori} on any of these parameters, comparing the estimates for ring features found in both C2685213 and C2685215 allows typical uncertainties on these parameters to be estimated. The differences in the estimated widths of the peaks from the two images are comparatively large, so estimates of those parameters are probably the least reliable. By contrast, the positions of the features generally agree to within a couple of kilometers (the exception being the relatively broad ring feature found near 47,860 km).  Furthermore, both the peak brightness and the NEW estimates have average fractional differences of around 25\%, so the uncertainties in these parameters are of that order.

\begin{table*}[p]
\caption{Parameters for Ringlets in Voyager NAC images}
\label{params}
\resizebox{6.5in}{!}{\begin{tabular} {c c c c c c c}\hline
Ringlet & Image Name & Observed  &  Predicted & FWHM  & Peak I/F  & NEW \\
             &                   &  Radius (km) & Radius$^*$ (km) & (km)& (10$^{-3}$) & (m) \\ \hline
6           & C2685204 & 41820 & 41820 & 6.4 & 0.098 & 0.630 \\
 U41.98 & C2685204 & 41983 &            &13.0 & 0.065 & 0.816 \\
 5          & C2685204 & 42210& 42211 & 12.0 & 0.096 & 0.976 \\
 4          & C2685204 & 42523 & 42526 & 7.9 & 0.107 & 0.791 \\
 U42.83? & C2685204 & 42830 &          & 6.9 & 0.065 & 0.335 \\
 U43.00? & C2685204 & 43001 &         & 9.6 & 0.098 & 0.974 \\ \hline
 $\alpha$ & C2685208 & 44696 & 44695 &  6.7 & 0.131 & 1.026 \\
 $\beta$ & C2685208 & 45651 & 45652 & 6.8 & 0.159 & 1.118 \\
 U45.73?  & C2685208 & 45727 &         &6.6 & 0.078 & 0.549 \\ \hline
U47.27 & C2685213 & 47270 &             &8.0 & 0.100 & 0.740 \\
U47.38 & C2685213 & 47383 &            & 11.8 & 0.079 & 0.936 \\
U47.38 & C2685215 & 47384 &           & 12.8 & 0.082 & 1.000 \\
$\gamma$ & C2685213 & 47627 &47627 & 9.2 & 0.119 & 1.069 \\
$\gamma$ & C2685215 & 47627 & 47627 & 4.7 & 0.117 & 0.627 \\
U47.86 & C2685213 & 47862 &           & 28.6 & 0.104 & 1.803 \\
U47.86 & C2685215 & 47855 &           & 34.7 & 0.072 &  1.627 \\
U48.06 & C2685213 & 48060 &           &10.8 & 0.114 & 1.299 \\
U48.06 & C2685215 & 48060 &         & 7.9 & 0.140 & 1.083 \\
U48.14 & C2685213 & 48139 &         & 9.4 & 0.101 & 0.885 \\
U48.14 & C2685215 & 48139 &          & 4.8 & 0.119 & 0.604 \\
U48.23 & C2685213 & 48226 &          & 9.5 & 0.314 & 3.242 \\
U48.23 & C2685215 & 48225 &          & 5.8 & 0.448 & 2.722 \\
$\delta$ & C2685213 & 48301 &48303 & 16.1 & 0.101 & 1.129 \\
$\delta$ & C2685215 & 48299 & 48303 & 5.3 & 0.133 & 0.768 \\ \hline
$\epsilon$ & C2685224 & 51049 & 51043 & 23.2 & 0.175 & 3.730 \\ \hline
\end{tabular}}

$^*$  Predicted radii of named features are from the most recent models of these rings \citep{French20}.

\end{table*}

For the sake of simplicity, first consider the C2685224 data shown in Figure~\ref{e}. This profile only contains an isolated peak corresponding to the $\epsilon$ ring, which falls near the inner edge of the profile. Note that due to the substantial eccentricity of this ring, its predicted location is noticeably different between the NAC and WAC profiles. In this case, the width and peak height of the ringlet is similar between the two profiles, which is consistent with this ringlet being relatively broad.

For all of the other profiles, we find not only brightness peaks that can be attributed to the named rings, but also signals that appear to represent comparably narrow unnamed dusty rings. Some of these peaks are clearly seen in both the NAC and WAC profiles and so their existence are relatively secure. However, a few features are only marginally detectable in the NAC profiles. Giving each of these features a unique name at this point would be impractical, so we will instead designate them by the letter U followed by a number corresponding to their location in thousands of kilometers. In addition, we will refer to more marginal features as ``ring candidates'' and include a ? in their designation. Note that the distinction between secure and marginal features is based on visual inspection of the NAC profiles rather than any quantitative signal-to-noise metrics because the marginal features occur in the parts of the profiles most strongly affected by instrumental artifacts.

Data from the C2685204 image are shown in Figure~\ref{456}. There are clear peaks associated with the 6, 5 and 4 rings, with the width of the 6 ring being noticeably less than that of the 5 ring, and the 4 ring having an asymmetric profile with a sharp inner edge and a more diffuse outer boundary. However, there is also a clear peak in the profile around 41,980 km in both the NAC and WAC profiles. The high-resolution NAC profile shows this ringlet is comparable in width to the 5 ring and is comparable in peak brightness and NEW to the 4, 5, and 6 rings. We designate this ringlet as U41.98. Exterior to Ring 4, the NAC profile overlaps the locations of two more peaks in the WAC profile around 42,830 and 43,000 km. While there do appear to be narrow peaks in the NAC profile, the signal-to-noise on these peaks are poor due to being near the edge of the frame, where brightness variations due to instrumental artifacts are more prominent, and so we cannot robustly confirm the existence of these features. In fact, there are multiple peaks associated with both of these features that most likely represent variations in the background profile, which further complicate quantifying these features. We will therefore denote both these features as potential ring candidates U42.83? and U43.00?

The C2685208 data in Figure~\ref{ab} shows two clear peaks  at the expected locations of the $\alpha$ and $\beta$ rings. Beyond that, there are no clear additional examples of narrow ring features. The one candidate is the peak at 45,730 km which is clear in the WAC profile but marginal in the NAC profile (Note the periodic brightness modulations that occur in this region are likely due to instrumental artifacts near the edge of the frame). We therefore designate this feature as ring candidate U45.73?.

The most complex profiles are those from images C2685213 and C2685215, which are shown in Figure~\ref{gd}. In both profiles, we see peaks at the expected locations of both the $\gamma$ and $\delta$ rings. However, neither the C2685213 NAC profile nor the WAC profile shows a peak at the expected location of the $\eta$ ring\footnote{Note that the location of this ringlet only varies by a couple of kilometers within the various occultations \citep{Chancia17}.}. Instead, there is a broad hump just exterior to that ring's location, which is clear in the WAC profile and might be present in the NAC profile. This could correspond to the 50-km wide low-optical depth extension to this ringlet first clearly seen in the Voyager radio occultations \citep{Gresh89, French91, dePater13}. There is also a distinct narrow peak at 47,270 km in the NAC profile, which we designate U47.27. This feature is also present in the WAC profile, but is not well separated from the broader feature closer to the $\eta$ ring. Note the peak in the NAC profile is roughly 100 km exterior to the expected location of the $\eta$ ring, and so is well beyond the edge of the low-optical-depth material seen in the occultations.

Continuing outwards, we find another, somewhat broader peak in both NAC profiles around 47,380 km that matches a distinct peak in the WAC profile. We designate this feature U47.38. Both profiles also show a peak at the expected location of the $\gamma$ ring, and further out around 47,800 km we see a general depression in the brightness that could reflect the dip in brightness seen in the WAC profile. However, the more prominent feature is a rather broad peak around 47,860 km, which corresponds to the similarly distinct peak in the WAC profile. We designate this feature U47.86. 

Between 48,000 and 48,300 km there are three narrow, roughly evenly-spaced ringlets that are comparable in width to the $\gamma$ and $\delta$ rings, which we designate U48.06, U48.14 and U48.23. The last of these is notable in that it is the brightest feature in the NAC images. Finally, we can note that while the peak associated with the $\delta$ ring itself is narrow, it also shows an asymmetric profile with a sharp outer edge and a more gradual inner edge. The slower drop on the inner edge could be due to the $\sim$10 km wide region of lower optical depth seen in Voyager radio occultations of this ring \citep{Gresh89, French91}.

\section{Discussion}
\label{discussion}

The NAC images reveal  that at least half a dozen of the ringlets seen in the high-phase Voyager observations have widths of order 10 km, which is comparable to the observed widths of the named dense ringlets in this region. This means that the $\lambda$ ring is not the only narrow dusty ringlet orbiting Uranus, instead it is the brightest member of an entire family of narrow dusty ring features. 

The small particles that dominate these features are sensitive to a range of non-gravitational forces that can radially transport material, including plasma or atmospheric drag and the Poynting-Robertson effect \citep{Grun84, Mignard84, Esposito91, French91, BHS01}. Thus there must be some outside force responsible for trapping dusty material at specific locations. In principle, each dusty ringlet could consist of material released from larger source bodies, but this then raises the natural question of why those source bodies are located at these locations. 

Most efforts to explain the distribution of dusty ringlets have been part of a general effort to explain the locations of the Uranian rings in terms of mean-motion resonances with various satellites  \citep{DG77, GN77, PG87a, PG87b, MT90, French91, Esposito91, Showalter95}.   However, thus far only a very  limited number of Uranian ring features can be explained by satellite resonances.

The best example of satellite resonances sculpting material in the Uranian ring system is the $\epsilon$ ring, whose inner and outer edges fall near resonances with the moons Cordelia and Ophelia, and the edges of this ring may even show the expected perturbations from these moons \citep{FN95}.  For all the other rings the connections with satellite resonances are less clear-cut. The $\eta$-ring's shape is clearly affected by the nearby 3:2 resonance with Cressida \citep{Chancia17} , but the resonance falls 5 km interior to the actual location of that ring so it is not clear whether the resonance plays any direct role in confining it. On the other hand, the brightness of the $\gamma$ ring is affected by a 6:5 resonance with Ophelia \citep{Showalter11}, but the overall shape of the ring is dominated by patterns with different azimuthal wavenumbers \citep{French91}, so it is still unclear how the resonance is influencing the dynamics of this ring. Finally, the $\delta$ ring lies near the 23:22 resonance with Cordelia, but its shape doesn't seem to be strongly influenced by this resonance \citep{French91}.  Furthermore, the 4,5,6 $\alpha$, $\beta$ and $\lambda$ rings are not particularly close to any resonances with Uranus' known satellites.  

Efforts have even been made to identify patterns in the locations of narrow ring features that might reflect resonances with unseen moons. \citet{MT90} found a few locations where a moon could be located that would explain 2 or 3 dense ring features, \citet{Horn88} attributed a feature in the $\delta$ ring to an unseen satellite,  and \citet{Showalter95} identified longitudinal variations in the $\lambda$-ring's brightness and location that were consistent with the expected perturbations from a satellite orbiting at 56,479 km. However, thus far none of these hypothetical moons have been detected, nor do they have resonances that match the locations of most of the other narrow features in the ring system.

\begin{figure}
\resizebox{3.3in}{!}{\includegraphics{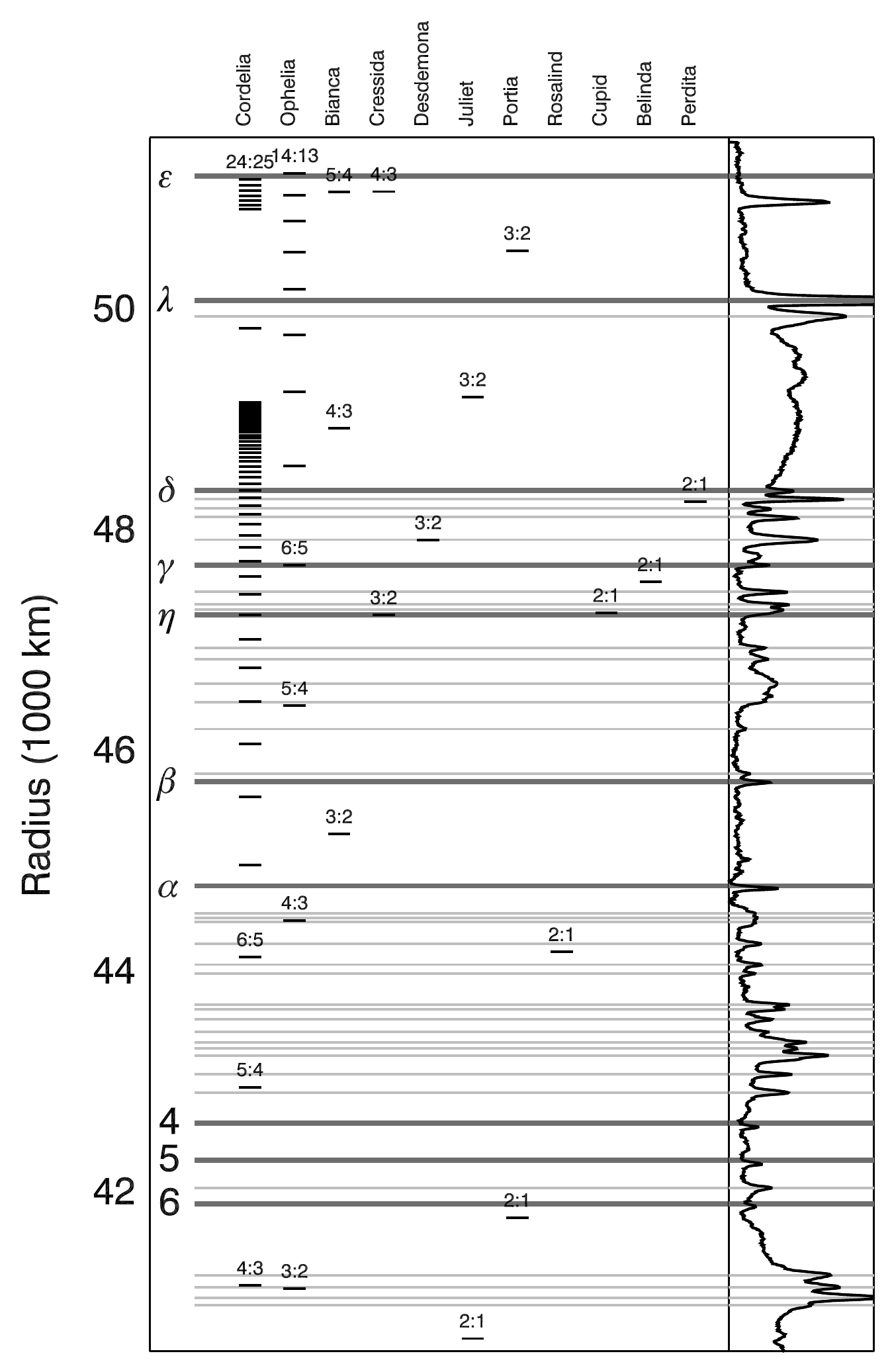}}
\caption{Comparisons between ring locations and the expected resonances with known moons \citep[updating a similar plot by ][]{MT90}. The dark grey bands indicate the semi-major axes of the named rings, while the light grey bands show the locations of dusty features found in the WAC profile at right. Note the locations of the named rings are sometimes offset due to the finite eccentricities of these features. Short dashes mark the predicted locations of first-order resonances with the known moons (For Cordelia, an isolated dash just interior to the $\lambda$ ring marks the semi-major axis of the moon itself and only mean-motion resonances greater than some finite distance from that location are marked). Note that the correlation between resonance locations and ring features is poor interior to 46,000 km.}
\label{res1}
\end{figure}

\begin{figure}
\resizebox{3.3in}{!}{\includegraphics{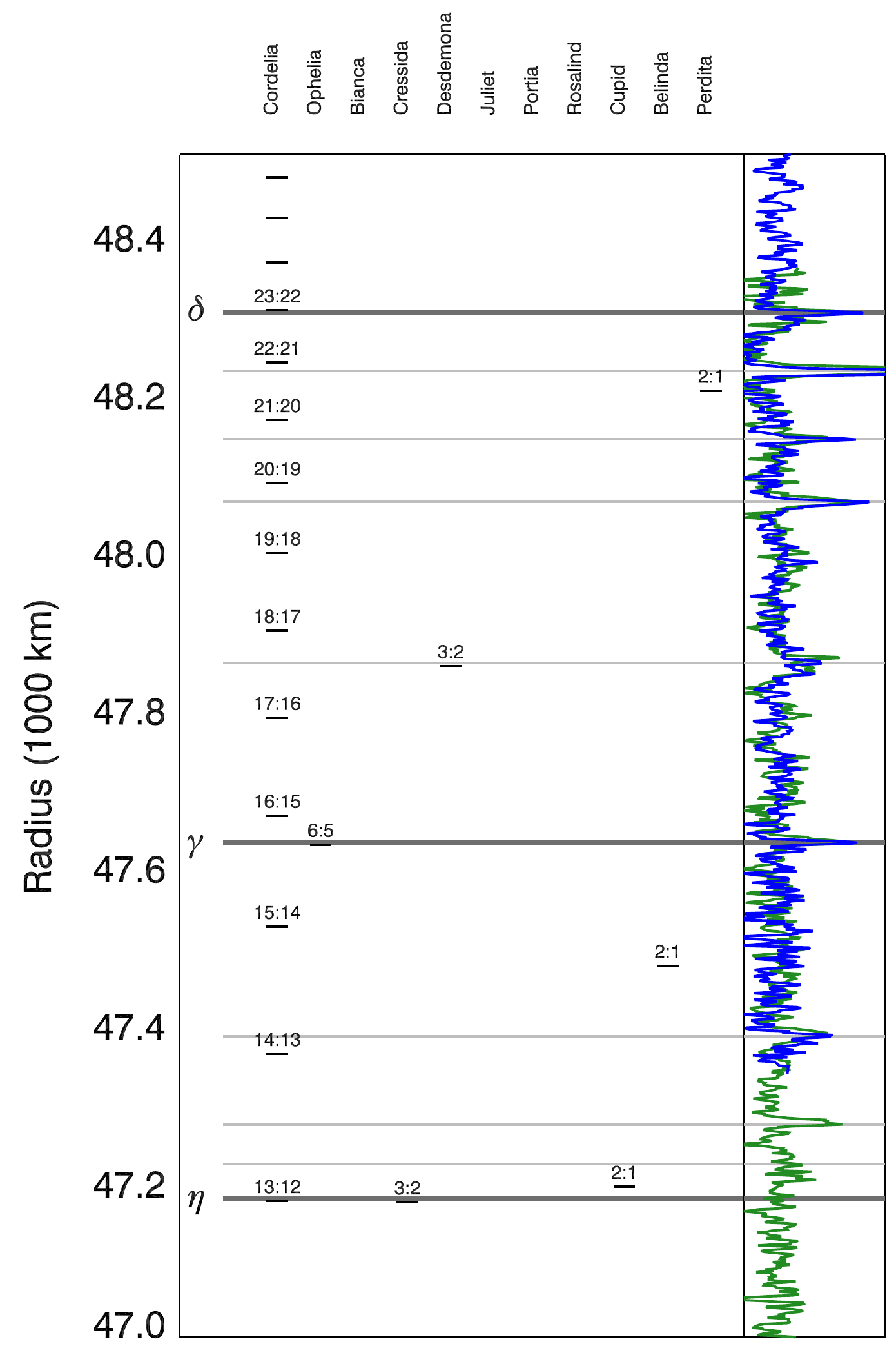}}
\caption{Comparisons between ring locations and the expected resonances with known moons for the region around the $\eta$, $\gamma$ and $\delta$ rings observed by the narrow-angle-camera images C2685213 and C2685215 (green and blue profiles at right). This plot shows that while the three named rings fall near satellite resonances, almost all of the dusty narrow rings are not located close to any satellite resonances. The one exception is U47.86, which is situated on top of the 3:2 resonance with Desdemona. Note in particular that none of the narrow dusty rings fall very close to the numerous Cordelia resonances.}
\vspace{1in}
\label{res2}
\end{figure}

The additional narrow dusty ring features identified in this study do not provide evidence to support the idea that satellite resonances are responsible for confining narrow rings around Uranus. Figure~\ref{res1} compares the locations of observed dense and dusty ring features with the predicted locations of first-order resonances with the Uranian moons. This plot clearly shows that interior to 46,000 km there are hardly any correlations between the locations of  either dense or dusty ring features and the satellite resonances, even if we consider the many narrow dusty features between the 4 and $\alpha$ rings.  Around the $\eta, \gamma$ and $\delta$ rings, however, one needs to take a closer look due to the higher density of resonances with Cordelia, and so Figure~\ref{res2} focuses on this particular region. Note that while the dense  $\eta, \gamma$ and $\delta$ rings are all found  close to satellite resonances \citep{PG87a, PG87b, Goldreich87, French88, French91, Chancia17}, the only dusty ringlet that lies comparably close to a resonance is U47.86, which lies on top of the 3:2 resonance with Desdemona. 

It is especially noteworthy that the three narrow dusty ringlets just interior to the $\delta$ ring are not well aligned with the Cordelia resonances, despite the rings and the Cordelia resonances having roughly similar spacings in this region. More specifically, the predicted locations of the 22:21, 21:20 and 20:19 resonances with Cordelia are 48236 km, 48163 km and 48083 km, which are 11 km, 24 km and 23 km exterior to the corresponding observed ring features. In principle, such offsets could arise because the ringlets are eccentric and so their observed locations in these images do not reflect their semi-major axes (note that the 6,5,4, $\alpha$ and $\beta$ rings all have eccentricities between 20 km and 80 km). 
However, it is unlikely that all three rings would be displaced in the same direction if we were observing three freely-precessing eccentric ringlets. In principle, distortions in the ringlets' locations due to the nearby Cordelia resonances could cause all three ringlets to be displaced in the same direction at certain locations, but in practice such alignments are unlikely to occur in this image because Cordelia is located at longitudes of 107.7$^\circ$ in C2685213 and 108.8$^\circ$ in C2685215, or over 140$^\circ$ from the observed ring longitudes. Furthermore, there is no clear reason why these four specific resonances would be the ones occupied by narrow rings. Hence at the moment there seems to be no good reason to attribute these particular narrow features to Cordelia resonances. 

These data therefore not only provide further evidence that the Uranian ring system has a high number of exceptionally narrow dusty rings, but also raises further questions about  how Uranus' narrow rings are confined. In all likelihood, a future spacecraft mission to Uranus capable of seeing and resolving these narrow dusty features and characterizing any longitudinal variations in their positions and brightness will be needed to fully answer this question. However, at the moment, it does not appear that satellite resonances alone will be able to explain the fine-scale structure of these dusty rings. Hence more exotic processes may need to be considered. 

In particular, it will probably be worth examining whether resonances with oscillations and asymmetries with Uranus' gravitational or electromagnetic fields could be sculpting these rings. Such possibilities are worth considering because structures have been found in Saturn's rings that are due to such planetary perturbations \citep{Chancia19, Hedman09, HN13, HN14, French16, French19, Hedman19}, the narrow rings around small bodies may be confined by spin-orbit resonances with asymmetries in the primary \citep{Sicardy19, OW19, Sumida20} and preliminary  calculations indicate that certain planetary oscillations could generate resonances in the vicinity of the $\delta$ and $\gamma$ rings \citep{Marley88}. 

\section*{Acknowledgements}

The authors thank M. Showalter, R. G. French and P.D. Nicholson for their helpful conversations regarding these images and their importance for understanding the structure of the Uranian rings.

\clearpage

%\bibliographystyle{icarus}
%\bibliography{uranus.bib} 

\end{document}